 \documentclass[nohyper,12pt,letterpaper]{JHEP3}
 \usepackage{graphicx}
 
 \def\bR{{\mathbb R}}

 \def\bD{{\mathbb D}} 
 \def\bZ{{\mathbb Z}}

 \newcommand{\vev}[1]{{\left< {#1} \right>}}

\title{Exact momentum fluctuations of an accelerated quark in ${\cal N}=4$ super Yang-Mills}

\author{Bartomeu Fiol, Blai Garolera and Gen\'is Torrents  \\

Departament de F{\'\i}sica Fonamental i \\Institut de Ci{\`e}ncies del Cosmos, 

Universitat de Barcelona,

Mart{\'\i}\ i Franqu{\`e}s 1, 08028 Barcelona, Catalonia, Spain \\

\email{bfiol@ub.edu, bgarolera@ffn.ub.es, genistv@icc.ub.edu }}

\abstract{In this work we consider a heavy quark moving with constant proper acceleration in the vacuum of any four dimensional conformal field theory. We argue that the two-point function of its momentum fluctuations is exactly captured by the Bremsstrahlung function that gives the total radiated power. For the particular case of ${\cal N}=4$ SU(N) SYM this function is exactly known, so in this case we obtain an explicit expression for the momentum diffusion coefficient, and check that various limits of this result are reproduced by probe computations in $AdS_5$. Finally, we evaluate this transport coefficient for a heavy quark accelerated in the vacuum of ${\cal N}=4$ SU(3) SYM, and comment on possible lessons of our results for the study of heavy quarks traversing the super Yang-Mills plasma.}  

\begin{document}
\section{Introduction}
One of the possible ways to study gauge theories is to introduce heavy external probes, following prescribed trajectories. These external probes can transform under various representations of the gauge group, and can also be coupled to additional fields, besides the gauge potential. A common tool to
implement this idea is the use of Wilson loops, where the contour of the loop is given by the world-line of the probe. Wilson loops are among the most interesting operators in a gauge theory, but in general computing their expectation value or their correlation functions with other operators is prohibitively difficult. On the other hand, for gauge theories with additional symmetries ({\it e.g.} conformal symmetry and/or supersymmetry) and for particular contours, a variety of techniques allows to prove exact relations among various correlators involving line operators, and sometimes also evaluate exactly these quantities \cite{Erickson:2000af, Drukker:2000rr,Pestun:2007rz, Pestun:2009nn}.

In this work we will be concerned with external probes coupled to a four dimensional conformal field theory (CFT), following either a static or a hyperbolic trajectory in vacuum. The probes can transform in an arbitrary representation of the gauge group, and when we consider probes transforming in the fundamental representation, we will often refer to them as quarks. We will extend recent work \cite{Correa:2012at, Fiol:2012sg} that provides exact relations among various observables related to these probes. In the particular case of ${\cal N}=4$ U(N) or SU(N) SYM, these exact relations will allow us to compute explicitly the momentum diffusion coefficient of an accelerated quark in vacuum, a transport coefficient than until now was only known in the limit of large N and large 't Hooft coupling $\lambda=g_{YM}^2N$.

The first observable that appears in our discussion is the energy emitted by a moving quark in accelerated motion, which for small velocities can be written as a Larmor-type formula
\begin{equation}
\Delta E=2\pi B(\lambda,N)\int dt \left(\dot v\right )^2
\label{bremdefi}
\end{equation}
where $B(\lambda,N)$ is a dimensionless function independent of the kinematics that was dubbed the Bremsstrahlung function in \cite{Correa:2012at}. 

One can also consider inserting operators on the world-line of the probe \cite{Polyakov:2000ti}. These operators localized on the world-line are not gauge singlets; nevertheless, their correlation functions evaluated on the world-line are gauge invariant. If the world-line is a straight line, it preserves a $SL(2,\bR)\times SO(3)$ subgroup of the original group \cite{Kapustin:2005py,Drukker:2005af}\footnote {This is the common group preserved by any line defect in any CFT. For particular CFTs with bigger symmetry groups, the preserved group might be much larger.}, and world-line operators can be classified according to representations of $SL(2,\bR)\times SO(3)$.
Among them, the so called displacement operators $\bD_i(t)$ $i=1,2,3$ \cite{Correa:2012at} will play a prominent role in this work. These are operators defined for any line defect in any conformal field theory, that couple to small deviations of the world line, orthogonal to it. They form a $SO(3)$ triplet and their scaling dimension $\Delta=2$ is protected for all values of the coupling, so their two-point function evaluated on a static world-line has the form
\begin{equation}
\vev{\vev{\bD_i(t)\bD_j(0)}}=\tilde \gamma(\lambda,N)\frac{\delta_{ij}}{t^4}
\label{twopoint} 
\end{equation}
where again $\tilde \gamma(\lambda,N)$ is a dimensionless coefficient and the double ket denotes evaluation on the world-line (see below for a precise definition). Physically, we will interpret correlators of these displacement operators as giving momentum fluctuations of the probe, an interpretation that has appeared before in the literature \cite{CasalderreySolana:2006rq, Gubser:2006nz} although not in this language. 

A crucial point for what follows is that the two coefficients in (\ref{bremdefi}) and (\ref{twopoint}) are exactly related by \cite{Correa:2012at}

\begin{equation}
\tilde \gamma =12 B
\label{therela}
\end{equation}
for any CFT and any straight line defect operator (Wilson loop, 't Hooft loop,$\dots$). This relation is claimed to be exact, valid for any value of the coupling constant, any gauge group and any representation of the gauge group. While it is important to appreciate that this Bremsstrahlung function appears in various observables related to probes coupled to CFTs, it is also important to actually compute it for different line operators in different interacting CFTs. Currently, this has only been done for a probe in the fundamental representation of ${\cal N}=4$ U(N) or SU(N) SYM, for which the Bremsstrahlung function $B(\lambda,N)$ was recently computed in \cite{Correa:2012at, Fiol:2012sg} and is given by

\begin{equation}
B_{U(N)}(\lambda,N)=\frac{\lambda}{16\pi^2 N}\frac{L_{N-1}^2\left(-\frac{\lambda}{4N}\right)+L_{N-2}^2\left(-\frac{\lambda}{4N}\right)}{L_{N-1}^1\left(-\frac{\lambda}{4N}\right)}
\label{brems}
\end{equation}
where the $L_n^\alpha$ are generalized Laguerre polynomials. It is worth emphasizing that this formula is valid for any value of $\lambda$ and $N$, and its derivation ultimately relies on localization techniques. In various limits, it can be checked using the AdS/CFT correspondence \cite{Mikhailov:2003er, Fiol:2011zg} or Bethe ansatz techniques \cite{Correa:2012nk}. To obtain the result for the $SU(N)$ theory, we have to subtract the $U(1)$ contribution \cite{Correa:2012at}

$$
B_{SU(N)}=B_{U(N)}-\frac{\lambda}{16 \pi^2N^2}
$$

The main observation of this paper is that the coefficient $\tilde \gamma$ in (\ref{twopoint}) also controls the two-point function of displacement operators when the probe is undergoing motion with constant proper acceleration $a=1/R$, since this hyperbolic world-line is related to the static one by a special conformal transformation. As it is well-known, a particle moving with constant proper acceleration in vacuum will feel an Unruh temperature $T=a/2\pi$ \cite{Unruh:1976db}. The thermal bath felt by the accelerated particle will cause momentum fluctuations, and these can be encoded in a particular transport coefficient, the momentum diffusion coefficient, defined as the zero frequency limit of the two-point function of displacement operators in momentum space,
$$
\kappa_{ij}\equiv \lim_{w\rightarrow 0} \int_{-\infty}^\infty d\tau e^{-iw\tau} \vev{\vev{\bD_i(\tau)\bD_j(0)}}
$$
where $\tau$ is the proper time of the accelerated probe. Since the hyperbolic trajectory still preserves a $SL(2,\bR)\times SO(3)$ subgroup of the original group \cite{Kapustin:2005py,Drukker:2005af}, by isotropy there is only a single transport coefficient as seen by the accelerated observer, $\kappa_{ij}=\kappa \delta_{ij}$, and a straightforward computation yields
\begin{equation}
\kappa = 16  \pi^3 B(\lambda,N) T^3
\label{ourrela}
\end{equation}
This is one of the main results of this paper; it relates the momentum diffusion coefficient of an accelerated heavy probe in the vacuum of a 4d CFT with
the corresponding Bremsstrahlung function, eq. (\ref{bremdefi}), and Unruh temperature. We claim that this result is exact for any 4d CFT, for any value of $\lambda$ and $N$ and for any representation of the gauge group. In the particular case of a heavy probe in the fundamental representation of ${\cal N}=4$ U(N) or SU(N) SYM, since $B(\lambda,N)$ is given exactly by (\ref{brems}), the relation (\ref{ourrela}) provides an explicit expression for the momentum diffusion coefficient. Furthermore, the result thus obtained can be subjected to a non-trivial check, since for ${\cal N}=4$ SYM , $\kappa$ has been computed in the large $\lambda$, planar limit by means of the AdS/CFT correspondence \cite{Xiao:2008nr,Caceres:2010rm}. Reassuringly, in the corresponding limit our result reduces to the previoulsy known one.

Having obtained the exact two-point function of momentum fluctuations of an accelerated heavy quark in the vacuum of ${\cal N}=4$ SU(N) SYM, it's tempting to ask whether we can use it to learn something about momentum fluctuations of a heavy quark in the midst of a finite temperature ${\cal N}=4$ SU(N) SYM plasma. This is a problem that has been extensively scrutinized in the context of the AdS/CFT correspondence \cite{Herzog:2006gh, Gubser:2006bz, CasalderreySolana:2006rq, CasalderreySolana:2007qw} (see \cite{CasalderreySolana:2011us} for reviews), as a possible model for the momentum fluctuations of a heavy quark traversing the quark-gluon plasma. In general, while an accelerated probe in vacuum registers a non-zero temperature, its detailed response differs from that of a probe in a thermal bath; in particular their respective retarded Green functions and momentum diffusion coefficients are different. Nevertheless, if we take the guess

$$
\frac{\kappa_{Unruh}^{exact}}{\kappa_{Unruh}^{SUGRA}}\approx \frac{\kappa_{thermal}^{exact}}{\kappa_{thermal}^{SUGRA}}
$$
for some range of values of $\lambda$ as a working hypothesis, we can estimate $\kappa_{thermal}^{exact}$ in that range of values of $\lambda$, since now the other three quantities in the relation above are known.

The plan of the paper is as follows. In section 2 we recall the definition of displacement operators, and we interpret their correlation functions as characterizing momentum fluctuations of the probe. We then go on to compute their exact two-point function for an accelerated infinitely heavy probe coupled to a CFT, and extract from it the momentum diffusion coefficient. In section 3, we use the AdS/CFT correspondence to verify the relation (\ref{therela}) in the particular case of ${\cal N}=4$ SU(N) SYM for static heavy probes in the fundamental and the symmetric representation. In section 4 we again use the AdS/CFT correspondence, now to compute the momentum diffusion coefficient of accelerated probes, in the fundamental and in the symmetric representations, and check that the results obtained are compatible with the exact result.  Finally, in section 5 we explicitly evaluate the momentum diffusion coefficient for
an accelerated quark coupled to ${\cal N}=4$ SU(3) SYM. We then evaluate the error introduced when one uses the supergravity expression instead of the exact one, and end by commenting on possible implications for the study of heavy quarks in a thermal bath.

\section{Momentum fluctuations and displacement operators}
Consider a heavy probe coupled to a four dimensional conformal field theory. This probe transforms in some representation of the gauge group, and perhaps it is also coupled to additional fields, as it is the case for $1/2$ BPS probes of ${\cal N}=4$ SYM \cite{Rey:1998ik, Maldacena:1998im}. Since we are considering a heavy probe, we will represent it by the corresponding Wilson line. In this section we will first recall the definition of certain operators inserted along the world-line, the displacement operators, and argue that their physical interpretation is that of momentum fluctuations due to the coupling of the probe to the quantum fields. We will then focus on the two-point function of such displacement operators for static and accelerated world-lines.

To define the displacement operators \cite{Correa:2012at}, consider a given Wilson loop, parameterized by $t$ and perform an infinitesimal deformation of the contour $\delta x^{\mu}(t)$, orthogonal to the contour $\delta \vec x(t)\cdot \dot{\vec x}(t) =0$. This deformation defines a new contour, and the displacement operator $\bD_i(t)$ is defined as the functional derivative of the Wilson loop with respect to this displacement \cite{Drukker:2011za}. In particular, the infinitesimal change can be written as
\begin{equation}
\delta W=P\int dt \delta x^j(t) \bD_j(t) W
\label{displa}
\end{equation}
These operators, and in general other local operators inserted on the world-line, are not gauge invariant. Nevertheless, their n-point functions, evaluated over the world-line are gauge invariant, {\it e.g.}
$$
\vev{\vev {\bD_i(t_1)\bD_j(t_2)}}=\frac {<Tr[P\bD_i(t_1)e^{i\int_{t_2}^{t_1}A\cdot dx}\; \bD_j(t_2)e^{i\int_{t_1}^{t_2}A\cdot dx}]>}{<Tr[Pe^{i\oint A\cdot dx}]>}
$$

What is the physical interpretation of these operators? In general, when we introduce an external heavy probe, its classical trajectory is fixed, giving the contour of the corresponding Wilson line. At the quantum level, this trajectory will suffer fluctuations due to its coupling to quantum fields. By definition, these small deformations in the contour $\delta x^i(t)$ are coupled to the displacement operators, so we identify these operators as forces causing momentum fluctuations. This identification is valid for general quantum field theories (not just conformal ones), and it has appeared before in the literature \cite{CasalderreySolana:2006rq, Gubser:2006nz}. For instance, if we consider a charged particle coupled to a $U(1)$ Maxwell field and moving with 4-velocity $U^\mu$, the Lorentz force is $qF_{\mu \nu}U^\nu$ and the displacement operators are
$$
\bD_\mu=qF_{\mu \nu}U^\nu 
$$
Since $U^\mu\bD_\mu=0$, we see explicitly that displacement operators are transverse to the world-line. This is easily generalized to additional couplings to scalar fields.

When the gauge theory under consideration is conformal, there is more we can say about displacement operators. Let's start by considering the world-line corresponding to a static probe, a straight line parameterized by $t$. In any conformal field theory, any straight line defect (or for that matter, any circular defect in Euclidean signature) preserves a $SL(2,\bR)\times SO(3)$ symmetry group of the original conformal group \cite{Kapustin:2005py, Drukker:2005af}, so operators inserted on the world-line are classified by their $SL(2,\bR)\times SO(3)$ quantum numbers. In particular, displacement operators $\bD_i(t)$ form a $SO(3)$ triplet, and since $\delta x^i$ and $t$ have canonical dimension, we learn from eq. (\ref{displa}) that displacement operators have scaling dimension $\Delta=2$, and this dimension is protected against corrections. This fixes their two-point function evaluated on a straight line to be of the form
\begin{equation}
\vev{\vev{\bD_i(t)\bD_j(0)}}=\tilde \gamma \frac{\delta_{ij}}{t^4}
\label{straight}
\end{equation}
Let's now consider a heavy probe moving with constant proper acceleration $a=1/R$ in one dimension. It is a textbook result that the resulting trajectory is the branch of a hyperbola in spacetime. Furthermore, for a conformal field theory, this hyperbolic world-line can be obtained by applying a special conformal transformation to a static world-line (the Euclidean counterpart of this statement is that a special conformal transformation can bring a straight line into a circle). This hyperbolic world-line also preserves a $SL(2,\bR)\times SO(3)$ symmetry group of the original conformal group \cite{Kapustin:2005py, Drukker:2005af}. If we write the two-point function of displacement operators in terms of proper time $\tau$ of the heavy probe, we have
\begin{equation}
\vev{\vev{\bD_i(\tau)\bD_j(0)}}=\tilde \gamma \frac{\delta_{ij}}{16R^4 \sinh^4 \left(\frac{\tau}{2R}\right)}
\label{twopointst}
\end{equation}
where the coefficient $\tilde \gamma$ is the same as for the two-point function evaluated on a straight line, eq. (\ref{straight}). This is required so at very short times, when $\tau/2R \ll 1$, we recover the result for the straight line, eq. (\ref{straight}).

\begin{figure}[htb]
\centering
\includegraphics[width=100mm,height=60mm]{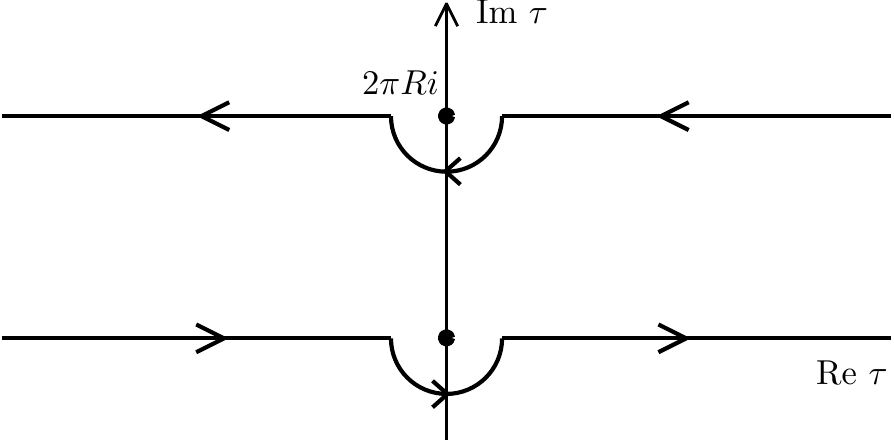}
\caption{Integration contour for the Fourier transform of the two-point function of displacement operators.}
\label{intecontour}
\end{figure}

On the other hand, the two-point functions (\ref{straight}) and (\ref{twopointst}) ought to reflect the very different physics felt by a static and an accelerated probe. In particular (\ref{twopointst}) captures the response of the accelerated probe to the non-zero Unruh temperature. To display this, we will now compute the Fourier transform of (\ref{twopointst}) and extract the corresponding transport coefficient. To compute the Fourier transform of (\ref{twopointst}), we notice that it presents poles in the $\tau$ complex plane whenever $\tau=2\pi i n R$, $n \in \bZ$. Using the same pole prescription as in \cite{Correa:2012at}, we follow \cite{Sciama:1981hr} and choose the integration contour displayed in figure (\ref{intecontour}). A straightforward computation then yields

$$
G(w)_{ij}=\int_{-\infty}^\infty d\tau e^{-iw\tau} \vev{\vev{\bD_i(\tau)\bD_j(0)}} = \tilde \gamma \delta_{ij}
 \int_{-\infty}^\infty d\tau \frac{e^{-iw\tau}}{16 R^4 \sinh^4 \left(\frac{\tau}{2R}\right)} 
=\tilde \gamma \delta_{ij} \frac{2\pi}{3!}\frac{\frac{w}{R^2}+w^3}{e^{2\pi w R}-1}
$$
This Green function displays a thermal behavior with temperature $T^{-1}=2\pi R$, i.e. the usual Unruh temperature. Note that this temperature depends only on the kinematics, not on dynamical aspects of the theory ({\it e.g.} it is coupling independent) \cite{Unruh:1983ac}.

We can take the zero frequency limit of this two-point function to obtain the momentum diffusion coefficients $\kappa_{ij}$. In fact, since the hyperbolic trajectory preserves a $SO(3)$ symmetry, there is a single transport coefficient $\kappa_{ij}=\kappa \delta_{ij}$ given by\footnote{This transport coefficient is usually obtained from the retarded Green function,
$$
\kappa=\lim_{w\rightarrow 0} -\frac{2T}{w}\hbox{ Im } G_R(w)
$$
Since for a static particle in a thermal bath, $G(w)=-\hbox {coth }\frac{w}{2T} \hbox{Im } G_R(w)$, in that case the two expressions are equivalent.
}
\begin{equation}
\kappa =\lim_{w\rightarrow 0} G(w)= 16  \pi^3 B(\lambda,N) T^3
\label{ourrelaag}
\end{equation}
This expression gives the momentum diffusion coefficient of an accelerated heavy probe in terms of the corresponding Bremsstrahlung function, eq. (\ref{bremdefi}) and the Unruh temperature $T$. It is valid for any four-dimensional CFT, and for any heavy probe. Notice that if we Fourier transform the two-point function evaluated on the straight line, eq. (\ref{straight}), we obtain that the Green function is proportional to $|w|^3$ \cite{Gubser:2006nz,Correa:2012at} so, as expected, the momentum diffusion coefficient defined as in (\ref{ourrelaag}) vanishes for a particle moving in vacuum with constant speed.

In the particular case of ${\cal N}=4$ SYM, the AdS/CFT correspondence provides the possibility of carrying out a non-trivial check of eq. (\ref{ourrelaag}). On the one hand, in the planar limit and for large $\lambda$, using the asymptotic value of $B \rightarrow \frac{\sqrt{\lambda}}{4\pi^2}$, eq. (\ref{ourrelaag}) reduces to
\begin{equation}
\kappa \rightarrow 4\pi \sqrt{\lambda} T^3
\label{ksugra}
\end{equation}
On the other hand, in this regime, one can use the AdS/CFT correspondence to compute this transport coefficient in an alternative fashion. The heavy probe is dual to a string reaching the boundary of $AdS_5$ at the hyperbolic world-line. Analysis of the fluctuations of this classical string solution allows to compute the relevant two-point function \cite{Xiao:2008nr,Caceres:2010rm} and from it extract the momentum diffusion coefficient \cite{Xiao:2008nr,Caceres:2010rm} (see also section 4), which precisely reproduces the result above, eq. (\ref{ksugra}). The dependence on $T$ was bound to agree, since is dictated by dimensional analysis, and the $\sqrt{\lambda}$ dependence is ubiquitous in AdS/CFT probe computations using fundamental strings (see {\it e.g.} \cite{Fiol:2012sg} for a discussion of this point), but the agreement on the numerical coefficient $4\pi$ in (\ref{ksugra}) is a non-trivial check.

\section{Static probes in AdS/CFT}
In this section we intend to verify the relation ($\ref{therela}$) for the particular case of ${\cal N}=4$ SYM by means of the AdS/CFT correspondence. To do so, we will compute separately $\tilde \gamma$ and $B$, and check that they are indeed related by $\tilde \gamma =12 B$. This relation ought to hold for a probe transforming in any representation of the gauge group, and we will perform this check for a heavy probe in the fundamental and in the k-symmetric representations. On the gravity side this corresponds to considering respectively a string and a D3-brane embedded in $AdS_5$ and reaching the boundary at a straight line. Performing the check for the k-symmetric representation has an interesting plus: while in principle the computation with the D3-brane can't be trusted in the limit when we set $k=1$ ({\it i.e.} we go back to the fundamental representation), by now there are a number of examples \cite{Drukker:2005kx,Giombi:2006de, Fiol:2011zg, Fiol:2012sg} where it is known that this procedure nevertheless correctly captures corrections in the large $\lambda$, large $N$ limit with fixed $\sqrt{\lambda}/N$. Given that we already know the exact expression of $B(\lambda,N)$ for this probe, eq. (\ref{brems}), we will be able to verify that this offers yet another example where a D3-brane probe computation correctly captures all order corrections to the leading large $\lambda$ large N result.

\subsection{Fluctuations of a static string in $AdS_5$}
The fluctuations of a static string in $AdS_5$ have been computed in many previous works \cite{Callan:1999ki, Forste:1999qn, Drukker:2000ep}, so we will be brief. We will work with the Nambu-Goto action in the static gauge, and will be concerned only with the bosonic fluctuations of the transverse coordinates in $AdS_5$, which we identify as dual to the world-line displacement operators. 

Start by writing $AdS_5$ in Poincar\'e coordinates
\begin{equation}
ds^2_{AdS_5}=\frac{L^2}{y^2}\left(dy^2-dt^2+d\vec x^2\right) 
\label{adspoincare}
\end{equation}
The relevant classical solution to the NG action is given by identifying the world-sheet coordinates with $(t,y)$. The induced world-sheet metric is $AdS_2$ with radius $L$. We now turn to the quadratic fluctuations around this solution, and focus on the fluctuations of the transverse coordinates in $AdS_5$, $x^i$, $i=1,2,3$. To make manifest the geometric content of these fluctuations, it is better to switch to $\phi^i=\frac{L}{y}x^i$. The Lagrangian density for quadratic fluctuations is then
\begin{equation}
{\cal L}=\frac{-1}{2\pi \alpha'}\left(-\frac{1}{2}\partial_t \vec \phi \; \partial_t \vec \phi +\frac{1}{2}\partial_y \vec \phi \; \partial_y \vec \phi +\frac{1}{y^2}(\vec \phi)^2\right)
\label{staticfluc}
\end{equation}
so the equation of motion for the fluctuations is 
$$
-\partial_t^2\phi^i+\partial_y^2\phi^i-\frac{2}{y^2}\phi^i=0
$$
from where we learn that the three transverse fluctuations in $AdS_5$ are massive $m^2=2/L^2$ scalars in the $AdS_2$ world-sheet \cite{Forste:1999qn,Drukker:2000ep}. Furthermore, it can also be seen that the five fluctuations of $S^5$ coordinates are massless \cite{Forste:1999qn,Drukker:2000ep}. The bosonic symmetries preserved by the classical string solution are $SL(2,\bR)\times SO(3)\times SO(5)$, which is the bosonic part of the supergroup $OSp(4^*|4)$ \cite{Nahm:1977tg}. Therefore, fluctuations should fall into representations of this supergroup, and indeed it is shown in \cite{Faraggi:2011bb} that together with the fermionic excitations they form an ultra-short multiplet.

These bosonic fluctuations are massive and massless scalars in the $AdS_2$ world-sheet, and according to the AdS/dCFT correspondence, ``holography acts twice" \cite{Karch:2000ct} and they source dual operators in the boundary of $AdS_2$, which is just the heavy quark world-line. The conformal dimensions $\Delta$ of these operators are determined by the usual relation
$$
2\Delta=d+\sqrt{d^2+4(mL)^2}
$$
In our case $d=1$, so the three fluctuations $\phi^i$ in $AdS_2$ with $m^2=2/L^2$ are dual to a   $SO(3)$ triplet of operators with $\Delta=2$: these are the displacement operators $\bD_i(t)$. Furthermore, the operators dual to the five massless $S^5$ fluctuations have $\Delta=1$  and are in the same supermultiplet as the displacement operators \cite{Correa:2012at}. We will not consider this second set of operators in the rest of the paper.

Our next objective is to compute the two-point function of displacement operators (\ref{twopoint}) in the regime of validity of SUGRA, i.e. the leading large $\sqrt{\lambda}$ large $N$ behavior of $\tilde \gamma (\lambda, N)$. This was essentially done in \cite{Gubser:2006nz}, with the minor difference that there the fluctuating fields were $x^i$ rather than $\phi^i$. After introducing the Fourier transform $x^i_F(w,y)$, the author of \cite{Gubser:2006nz} solved the corresponding equation and imposing purely outgoing boundary conditions, obtained the following Green function \cite{Gubser:2006nz}

$$
G(w)=\frac{L^2}{2\pi\alpha'}|w|^3 \Rightarrow G(t)=\frac{3\sqrt{\lambda}}{\pi^2}\frac{1}{t^4}
$$
from where we finally deduce
\begin{equation}
\tilde \gamma=\frac{3\sqrt{\lambda}}{\pi^2}
\label{gammasugra}
\end{equation}
To complete the check, we need the coefficient of energy loss by radiation, defined in eq. (\ref{bremdefi}).
The computation of $B$ for a heavy probe in this regime was first carried out by Mikhailov in a beautiful paper \cite{Mikhailov:2003er}, obtaining $B=\frac{\sqrt{\lambda}}{4\pi^2}$. Putting together these two results, we have verified $\tilde \gamma=12B$ to this order. 

\subsection{Fluctuations of a static D3-brane in $AdS_5$}
We will now check relation (\ref{therela}) for a heavy probe in the k-symmetric representation. To do so, we will consider a D3-brane dual to a static probe in $AdS_5$, with $k$ units of electric flux that 
encode the representation of the heavy probe. The relevant static D3-brane solution was found in \cite{Rey:1998ik, Drukker:2005kx}, but for our purposes it will be convenient to present it in the coordinates introduced in \cite{Yamaguchi:2006tq,Faraggi:2011bb}. First, write $AdS_5$ in the following coordinates
$$
ds^2_{AdS_5}=L^2\left(du^2+\cosh^2 u \frac{1}{r^2}\left(-dt^2+dr^2\right)+\sinh^2 u \left(d\theta^2+\sin^2\theta d\phi^2\right)\right)
$$
The D3-brane world-volume coordinates are $(t,r,\theta,\phi)$. The classical solution includes some non-trivial world-volume electric field
$$
\sinh u =\nu \hspace{1cm}F_{tr}=\frac{\sqrt{\lambda}}{2\pi}\frac{\sqrt{1+\nu^2}}{r^2}
$$
with\footnote{This combination was originally dubbed $\kappa$ in \cite{Drukker:2005kx} and subsequent works. To avoid any possible confusion with the momentum diffusion coefficient, in this work we change its name to $\nu$.}
\begin{equation}
\nu=\frac{k\sqrt{\lambda}}{4N}
\label{nudef}
\end{equation}
The induced metric is of the form $AdS_2\times S^2$, with radii $L\sqrt{1+\nu^2}$ and $L\nu$ respectively.
Consider now fluctuations of $u$, the $S^5$ coordinates and the Born-Infeld abelian gauge field $A_\mu$. The main advantage of the coordinates used here is that as shown in detail in \cite{Faraggi:2011bb} these sets of fluctuations decouple in these coordinates, so we can focus exclusively on fluctuations of $u$. Due to the presence of world-volume fluxes, the Lagrangian density for the fluctuating fields is not controlled by the induced world-volume metric, but by the $AdS_2\times S^2$ metric with both radii $L\nu$
$$
ds^2_{AdS_2\times S^2}={\cal G}_{ab}d\xi^ad\xi^b=\frac{L^2\nu^2}{r^2}\left(-dt^2+dr^2\right)+L^2\nu^2\left(d\theta^2+\sin^2 \theta d\phi^2\right)
$$
In particular, the Lagrangian density for fluctuations of $u$ is
$$
{\cal L}=-T_{D3}\frac{\sqrt{1+\nu^2}}{\nu}\sqrt{-|{\cal G}|}\left(\frac{1}{2}L^2{\cal G}^{ab}\partial _a u\partial _b u\right)
$$
Given that the D3-brane world-volume is of the form $AdS_2\times S^2$, the next step is to perform a KK reduction of these fields on the world-volume $S^2$ to end up with fields living purely on $AdS_2$. This produces an infinite tower of modes, but the only ones relevant for us are the $l=1$ triplet, since those are the ones sourcing the displacement operators. This KK reduction is discussed in detail in \cite{Faraggi:2011bb} (see their appendix C), and for the $l=1$ triplet of modes we are interested in, the resulting fluctuation Lagrangian is $k \sqrt{1+\nu^2}$ times the one computed with the string. Since the computation of the two-point function of displacement operators involves the kinetic term of the fluctuations, the upshot is that the $\tilde \gamma$ computed in this regime is $k \sqrt{1+\nu^2}$ times the one computed with the string in the previous subsection, eq. (\ref{gammasugra}), so
$$
\tilde \gamma(\lambda,N)=\frac{3 k\sqrt{\lambda}}{\pi^2}\sqrt{1+\frac{k^2\lambda}{16N^2}}
$$
To finish the check, we again need the coefficient $B$ in (\ref{bremdefi}) for this case. In \cite{Fiol:2011zg} the total radiated power of a heavy probe in the k-symmetric representation was computed using the AdS/CFT correspondence by means of a D3-brane, and the result found was
$$
B(\lambda,N)=\frac{k \sqrt{\lambda}}{4\pi^2}\sqrt{1+\frac{k^2\lambda}{16N^2}}
$$
Comparing these two results, this proves the $\tilde \gamma=12B$ relation for a static probe in k-symmetric representation, in the regime of validity of the D3-brane probe approximation. Furthermore, if we set $k=1$ in the previous result we can check \cite{Giombi:2006de, Fiol:2012sg} that the exact expression for $B(\lambda,N)$ reduces to the one above in the appropriate limit.

\section{Accelerated probes in AdS/CFT}
In this section we will consider accelerated heavy probes coupled to ${\cal N}=4$ SYM, in the context of the AdS/CFT correspondence. As in the previous section, the probes considered transform in the fundamental and the symmetric representations, so their gravity dual is given respectively by a string and a D3-brane, reaching the boundary at a hyperbola. We will compute the momentum diffusion coefficient in both cases, verifying that they reproduce in appropriate limits our exact result (\ref{ourrelaag}).

\subsection{Fluctuations of the hyperbolic string in $AdS_5$}
The dual of a heavy probe moving with constant proper acceleration is a string reaching the boundary of $AdS_5$ at a hyperbola, or at a circle in Euclidean signature. This type of world-sheet was first considered in \cite{Berenstein:1998ij}, see also \cite{Branding:2009fw}\footnote{Some subtleties associated to this world-sheet solution and its usual interpretation have been recently pointed out in \cite{Garcia:2012gw}, but since they concern the part of the world-sheet below the world-sheet horizon, they don't affect our discussion.}. The spectrum of fluctuations of this world-sheet was discussed in \cite{Drukker:2000ep}, and it is the same as for the straight line. This world-sheet and its fluctuations were used in \cite{Xiao:2008nr,Caceres:2010rm} to derive the momentum diffusion coefficient of this probe in the supergravity approximation. To do so, \cite{Xiao:2008nr,Caceres:2010rm} made a series of change of coordinates to the gravity background, to work in a frame where the probe is static. We will now show that it is possible to obtain that transport coefficient working with Rindler coordinates. We start by writing the $AdS_5$ metric in Poincar\'e patch with Rindler coordinates
$$
ds^2_{AdS_5}=\frac{L^2}{y^2}\left(dy^2+dr^2-r^2d\psi^2+dx_2^2+dx_3^2\right)
$$
We identify the world-sheet coordinates with $(\psi, y)$. The classical solution is then given by $r=\sqrt{R^2-y^2}$ \cite{Berenstein:1998ij}. We consider now fluctuations in the transverse directions $x^2,x^3$. The Lagrangian density for transverse fluctuations is
$$
{\cal L}_{fluc}=\frac{1}{2\pi \alpha'}\frac{L^2R}{y^2}\left(-\frac{1}{2}\frac{1}{R^2-y^2}\left(\partial_\psi x\right)^2+\frac{1}{2}\frac{R^2-y^2}{R^2}\left(\partial_y x\right)^2\right)
$$
As a check, near the boundary $(y\rightarrow 0)$, defining $\tau =R\psi$ we recover the fluctuation Lagrangian (\ref{staticfluc}), except for a global factor of $R$, since here we are integrating with respect to $\psi=\tau/R$. Defining $z=y/R$, the equation of motion for transverse fluctuations is
$$
-\partial_\psi^2 x+(1-z^2)^2\partial_z^2x-2\frac{1-z^2}{z}\partial_z x=0
$$
We separate variables $x(z,\psi)=e^{-iw\psi}x(z)$ (and keeping in mind that this $w$ is dimensionless, $w_\tau=w/R$), the solutions are
$$
x(z)=C_1(1-iwz)e^{iw \hbox{ arctanh }z}+C_2(1+iwz)e^{-iw \hbox{ arctanh }z}
$$
To compute the retarded Green function, we take the purely outgoing solution ($C_2=0)$ and following  \cite{Gubser:2006nz} obtain
$$
G_R(w)=\frac{-iw}{2\pi \alpha'}\frac{L^2}{R^2} +{\cal O}(w^3)
$$
where as in the static case  \cite{Gubser:2006nz} we dropped a $1/y$ term. This retarded Green function coincides with the one computed by Xiao \cite{Xiao:2008nr}, and from it one arrives at
$$
\kappa=4\pi \sqrt{\lambda}T^3
$$

\subsection{Fluctuations of a hyperbolic D3-brane in $AdS_5$}
We can also compute the momentum diffusion coefficient for an accelerated probe in the symmetric representation. The relevant D3-brane reaches the boundary at a circle in Euclidean signature, and was first discussed in \cite{Drukker:2005kx}. As in the previous section, it is convenient to present it in the coordinates introduced in \cite{Yamaguchi:2006tq,Faraggi:2011bb}, so we start by writing $AdS_5$ as
$$
ds^2=L^2\left(du^2+\cosh^2 u \left(d\zeta^2-\sinh^2 \zeta d\psi^2 \right)+\sinh ^2 u \; d\Omega_2 ^2\right)
$$
The D3-brane has world-volume coordinates $(\zeta,\psi,\theta,\phi)$ and the classical solution is
$$
\sinh u=\nu\hspace{1cm}F_{\zeta \psi}=\frac{\sqrt{\lambda}}{2\pi}\sqrt{1+\nu^2}\sinh \zeta
$$
with $\nu$ defined in eq. (\ref{nudef}). Consider now fluctuations for the world-volume fields $u$, the $S^5$ fields and the BI gauge field. As it was discussed in detail in \cite{Faraggi:2011bb} these fluctuations decouple, so we can focus on the fluctuations of $u$. To present the relevant fluctuation Lagrangian, define the metric
$$
{\cal G}_{ab}d\xi^a d\xi^b=L^2\nu^2(d\zeta^2-\sinh ^2\zeta d\psi^2)+L^2\nu^2 d\Omega_2^2
$$
This is the metric that controls the fluctuations of $u$ (and the gauge field)
$$
{\cal L}_{fluc}=T_{D3}L^4\sqrt{1+\nu^2}\nu^3\sinh \zeta \sin \theta
\left(\frac{1}{2}L^2{\cal G}^{ab}\partial_a u\partial_b u\right)
$$
As in the previous section, we now have to KK reduce this world-volume field $u$ on $S^2$, to obtain an infinite tower of 2d fields on the world-volume $AdS_2$. Again, the relevant modes are the $l=1$ triplet, and as it happened for the static probe, the resulting fluctuation Lagrangian is $k\sqrt{1+\nu^2}$ the one we would obtain for the fluctuations of a fundamental string in these coordinates. We then conclude that the resulting momentum diffusion coefficient is again $k\sqrt{1+\nu^2}$ times the one obtained for the fundamental string, so
$$
\kappa=4\pi k \sqrt{\lambda}\sqrt{1+\frac{k^2\lambda}{16N^2}} T^3
$$

\section{Lessons for the ${\cal N}=4$ super Yang-Mills plasma?}
In section 2 we have found the exact two-point function of momentum fluctuations in vacuum of an accelerated heavy quark coupled to a conformal field theory. As expected, this two-point function presents thermal behavior, and the question arises whether we can use our results to learn something about momentum fluctuations of a heavy probe immersed in a thermal bath of the same conformal theory, now at finite temperature. Besides its intrinsic interest, this question has broader relavance since it is expected that at finite temperature, conformal theories (even superconformal ones) share some properties with the high-temperature deconfined phase of confining gauge theories. More specifically, a particular CFT, ${\cal N}=4$ SYM at $T\neq 0$, has been used by means of the AdS/CFT correspondence to model the quark-gluon plasma experimentally observed at RHIC and at the LHC (see \cite{CasalderreySolana:2011us} for reviews). In particular, the momentum fluctuations of a heavy quark (either static or moving at constant velocity) in the quark-gluon plasma have been estimated by considering a dual trailing string in the background of a black Schwarzschild brane in an asympotically $AdS_5$ background \cite{Herzog:2006gh,CasalderreySolana:2006rq, Gubser:2006bz, CasalderreySolana:2007qw}.

The study of a heavy quark in a strongly coupled conformal plasma by means of the  AdS/CFT  correspondence is currently limited to the large $\lambda$, large N regime where supergravity is reliable (see \cite{Zhang:2012jd} for computation of the $1/\sqrt{\lambda}$ correction and some $\lambda^{-3/2}$ corrections to the jet quenching parameter in the ${\cal N}=4$ SYM plasma) and it currently seems extremely hard to perform such computations at finite $\lambda$ and $N$. For this reason, it would be very interesting if the study of an accelerated quark in the vacuum of a conformal field theory, which as we have seen can be tackled at finite $\lambda$ and $N$, can become an indirect route to the study of conformal $T\neq 0$ plasma. However, while a probe accelerated in vacuum and a static probe in a thermal bath experience a non-zero temperature, the details of their response are not identical ( see the review \cite{Crispino:2007eb} for a discussion on this point).  We can see this explicitly for the ${\cal N}=4$ SYM plasma, by comparing known expressions of the momentum diffusion coefficients in various regimes. Let's consider first the regime of weak coupling; the momentum diffusion coefficient of a heavy quark in a weakly coupled ${\cal N}=4$ SU(N) SYM plasma has been computed at leading and next-to-leading orders \cite{Chesler:2006gr}
$$
\kappa_{thermal}=\frac{\lambda^2 T^3}{6 \pi}\frac{N^2-1}{N^2}\left(\log \frac{1}{\sqrt{\lambda}}+c_1+c_2\sqrt{\lambda}+{\cal O}(\lambda)\right)
$$
with $c_{1,2}$ known coefficients (see the second reference in \cite{Chesler:2006gr}). This expression differs qualitatively from the weak coupling expansion of our result for the momentum diffusion coefficient for an accelerated quark
$$
\kappa_{Unruh}=\pi\lambda T^3\frac{N^2-1}{N^2}\left(1-\frac{\lambda}{24}+{\cal O}(\lambda^2)\right)
$$
Notice that $\kappa_{thermal}$ starts at $\lambda^2$ (versus the leading $\lambda$ in $\kappa_{Unruh}$) and furthermore presents a term logarithmic in $\lambda$, absent in $\kappa_{Unruh}$. These two features come from the non-trivial coupling dependence of the Debye mass in the thermal bath \cite{bellac}.

Let's move now to the regime where supergravity is reliable, {\it i.e.} large $\lambda$ and large $N$. As recalled in section 4, an accelerated probe in vacuum is dual to a string reaching the boundary of pure $AdS_5$ at a hyperbola, while a heavy probe in a thermal bath is represented by a string in the Schwarzschild- Anti de Sitter background, and the respective retarded Green functions are quantitatively different (see \cite{Hirayama:2010xi} for a discussion on this point). In particular, the momentum diffusion coefficient yields \cite{Herzog:2006gh,CasalderreySolana:2006rq}
$$
\kappa^{SUGRA}_{thermal}=\pi \sqrt{\lambda} T^3
$$
which is four times smaller than the supergravity result for the similar transport coefficient for a probe accelerated in vacuum, eq. (\ref{ksugra}),
$$
\kappa^{SUGRA}_{Unruh}=4\pi \sqrt{\lambda} T^3
$$
This difference might be surprising at first, since it can be argued that transport coefficients can be read from the world-sheet horizon \cite{Iqbal:2008by}, and the two classical world-sheet metrics ({\it i.e.} accelerated string in $AdS_5$ versus hanging/trailing string in Schwarzschild-$AdS_5$) while clearly different, have the same near-horizon metric, 1+1 Rindler space. However, the different change of variables used to write these near-horizon metrics imply different normalizations of the corresponding wavefunctions, giving rise to this factor of four discrepancy between the respective transport coefficients.

Keeping this difference in mind, we nevertheless propose to use our exact results to make an educated guess of the impact of using SUGRA instead of the exact result for computing the momentum diffusion coefficient of a static heavy quark, $\kappa_{thermal}$, in ${\cal N}=4$ SYM at finite temperature. To that end, we start by evaluating the difference between the SUGRA (large $\lambda$, large $N$) and the exact (finite $\lambda$, N=3) computations of the coefficient for the accelerated probe in vacuum.

The first ingredient we need in our computation is the Bremsstrahlung function (\ref{bremdefi}) for a heavy quark coupled to ${\cal N}=4$ SU(3) SYM. For $U(N)$ the Bremsstrahlung function is given in (\ref{brems}), and since the $SU(N)$ function is obtained  by subtracting the $U(1)$ contribution \cite{Correa:2012at}
$$
B_{SU(N)}=B_{U(N)}-\frac{\lambda}{16 \pi^2N^2}
$$
we obtain  
$$
B_{SU(3)}=\frac{1}{4\pi^2}\frac{\lambda}{18}\frac{\lambda^2+144\lambda+3456}{\lambda^2+72 \lambda+864}
$$
and using the relation derived in this paper, eq.  (\ref{ourrelaag}), we arrive at the following expression for the $SU(3)$ momentum diffusion coefficient, valid for any value of $\lambda$,
\begin{equation}
\kappa_{SU(3)}=4\pi \frac{\lambda}{18}\frac{\lambda^2+144\lambda+3456}{\lambda^2+72 \lambda+864}\; T^3
\label{ourkappa}
\end{equation}
Notice that both for small $\lambda$ and large $\lambda$ the coefficient grows linearly with $\lambda$. This is true for generic fixed $N$
$$
\kappa_{SU(N)}^{\lambda\ll 1}=\frac{N^2-1}{N^2}\pi \lambda T^3 \hspace{1cm} \kappa_{SU(N)}^{\lambda\gg 1}=\frac{N-1}{N^2}\pi \lambda T^3
$$
We now consider the quotient of the exact expression for this transport coefficient, eq. (\ref{ourkappa}), versus the result obtained in the supergravity limit, eq. (\ref{ksugra}), 
\begin{equation}
\hbox{Unruh}\hspace{1cm} \frac{\kappa^{EXACT}}{\kappa^{SUGRA}}=\frac{\sqrt{\lambda}}{18}\frac{\lambda^2+144\lambda+3456}{\lambda^2+72 \lambda+864}
\label{unruhratio}
\end{equation}
A first observation is that this ratio is a monotonously increasing function of $\lambda$ that doesn't go to one as $\lambda\rightarrow \infty$. The reason is that the denominator, obtained in the planar limit ($N\rightarrow \infty$), grows like $\sqrt{\lambda}$, while the numerator, obtained for $N=3$,
grows like $\lambda$. This ratio is smaller than one for small $\lambda$ and becomes larger than one for $\lambda \gtrsim 182.45$. As we discuss below, when modelling the quark-gluon plasma by ${\cal N}$=4 SYM the range of values considered for $\lambda$ is substantially below this point, so another observation is that in that range of values, the supergravity computation gives a value $\kappa^{SUGRA}$ which is larger than $\kappa^{EXACT}$. To be more quantitative, we will zoom in the range of values of $\lambda$ that have been considered when modelling the QCD quark-gluon plasma by ${\cal N}=4$ SYM. Given the differences among these two theories, there are inherent ambiguities in choosing the parameters of ${\cal N}=4$ SYM that might best model the real world QCD plasma. A first choice \cite{Liu:2006ug} 
is to take 
$$
\hbox{``obvious" scheme:} \hspace{1cm}T_{{\cal N}=4}=T_{QCD} \hspace{1cm} g^2_{YM}N=12\pi\alpha_s=6\pi
$$
where in the last equation the value $\alpha_s=0.5$ was taken. A second choice made in the literature \cite{Gubser:2006qh, Gubser:2006nz} tries to ameliorate the impact of the obvious difference that ${\cal N}=4$ SU(3) SYM has more degrees of freedom than QCD. The main idea is to compare the theories at fixed value of the energy density, rather than temperature. This results in the following identification
$$
\hbox{``alternative" scheme:}\hspace{1cm} 3^{1/4}T_{{\cal N}=4}=T_{QCD} \hspace{1cm} g^2_{YM}N=5.5
$$
where the value $\lambda=5.5$ is the central value derived from this analysis. This scheme has its own limitations, and the only lesson we want to take from it is that the range of values $\lambda \in [5.5,6 \pi]$ has been considered when modelling the quark-gluon plasma by the ${\cal N}=4$ SYM plasma. Having fixed the range of values for $\lambda$ we will be zooming in, we can now determine the impact of using supergravity to compute the momentum diffusion coefficient instead of using the exact result: in this range, the ratio (\ref{unruhratio}) increases from 0.43 to 0.61, see figure (2). Roughly speaking, in this range of values for the 't Hooft coupling, supergravity gives an answer for this transport coefficient about twice the exact result.

\begin{figure}[htb]
\centering
\includegraphics[width=70mm,height=50mm]{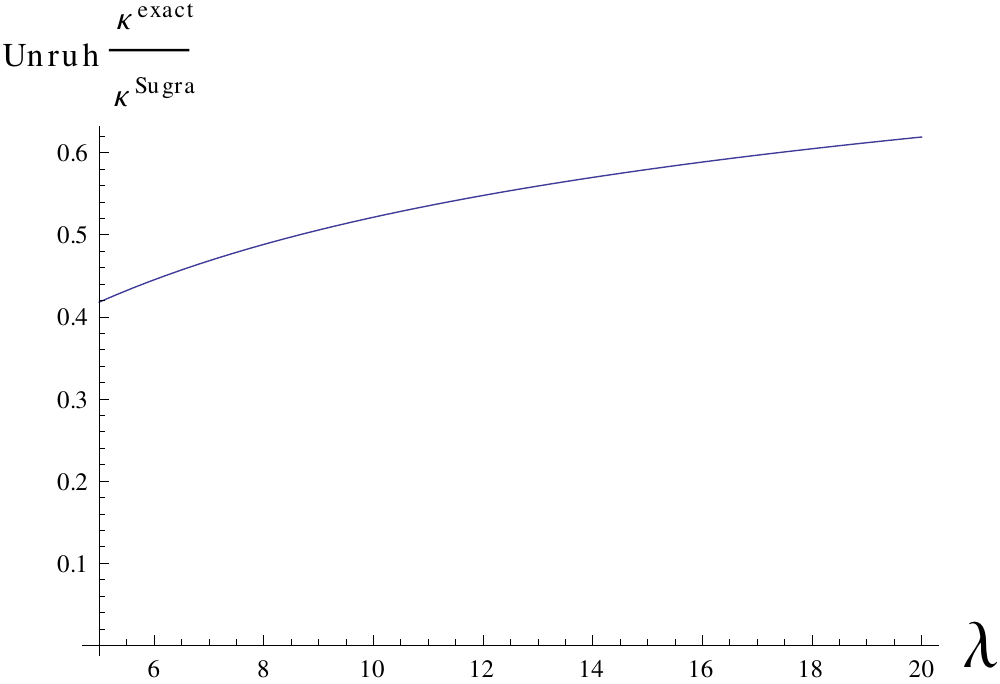}
\caption{The relation between the exact momentum diffusion coeffient and the supergravity approximation for an accelerated quark in vacuum. The range of 
$\lambda$ displayed corresponds to the one considered when modelling the quark-gluon plasma.}
\end{figure}

Up until here, we were on firm ground, comparing the result of two computations for the momentum diffusion coefficient of an accelerated quark in the vacuum of ${\cal N}=4$ SYM. Having used the word ``exact" or variations exactly thirty-three times so far in this paper, we will end it by indulging in far less precise statements. We have seen that the perturbative expressions (fixed $N$, $g_{YM}^2\ll 1$) for the momentum diffusion coefficients in a thermal bath and in the Unruh effect differ qualitatively, while the corresponding expressions in the supergravity regime (large $\lambda$, large $N$), share the same parametric behavior, but not the overall numerical coefficient. It is then clear that we are not in a position to estimate $\kappa_{thermal}$ for ${\cal N}=4$ SYM for abitrary $\lambda, N$. A more modest goal is to estimate it in the range of values singled out above, that appear when modelling the QCD quark-gluon plasma. If we consider a path in the $(\lambda,N)$ plane from the range of values considered above to the region of validity of supergravity ({\it i.e.} now we don't keep $N$ fixed) the ratio $\kappa_{Unruh}/\kappa_{Unruh}^{SUGRA}$ will uneventfully evolve from the value found above, about 1/2, to 1. In order to proceed, we are going to assume that roughly the same is true for $\kappa_{thermal}$ so along that path
$$
\frac{\kappa_{Unruh}^{exact}}{\kappa_{Unruh}^{SUGRA}}\approx \frac{\kappa_{thermal}^{exact}}{\kappa_{thermal}^{SUGRA}}
$$
If this assumption is true, it means that the supergravity computations \cite{CasalderreySolana:2006rq, Herzog:2006gh, Gubser:2006bz} for $\kappa_{thermal}$ give an answer $\kappa_{thermal}^{SUGRA}$ that is about twice the exact one. While we currently lack solid arguments to substantiate this speculation, let's end by noting that if true, it would in turn imply that the diffusion constant $D=2T^2/\kappa$ for the ${\cal N}=4$ SYM plasma would be about twice the one obtained in supergravity, pushing it in the right direction to match the range of values suggested by RHIC \cite{Francis:2011gc}.

\section{Acknowledgements} 
We would like to thank Jorge Casalderrey-Solana for many helpful conversations and explanations. We have also benefited from conversations with Nadav Drukker, Roberto Emparan and David Mateos. BG would like to thank Diego Correa and the Departamento de F\'isica at the Universidad Nacional de la Plata for hospitality during the course of this work. The research of BF is supported by MEC FPA2010-20807-C02-02, CPAN CSD2007-00042, within the Consolider-Ingenio2010 program, and AGAUR 2009SGR00168. BG is supported by an ICC scholarship and by MEC FPA2010-20807-C02-02, and GT by an FI scholarship by the Generalitat the Catalunya.

\end{document}